\documentclass[showkeys,floatfix,eqsecnum,prb]{revtex4}
\usepackage{graphicx}
\usepackage{dcolumn}
\usepackage{bm}
\usepackage{hyperref}

\begin{document}

\title{Monte Carlo simulation of boson lattices}
\author{Vesa Apaja}
\affiliation{Institut f{\"u}r Theoretische Physik, Johannes-Kepler Universit\"at, A-4040 Linz, Austria}
\author{Olav F. Sylju{\aa}sen}
\affiliation{Nordita, Blegdamsvej 17, DK-2100, Copenhagen \O, Denmark}

\begin{abstract}
Boson lattices are theoretically well described by the Hubbard
model. The basic model and its variants can be effectively simulated
using Monte Carlo techniques. We describe two newly developed
approaches, the Stochastic Series Expansion (SSE) with directed loop
updates and continuous--time Diffusion Monte Carlo (CTDMC).  SSE is a
formulation of the finite temperature partition function as a
stochastic sampling over product terms.  Directed loops is a general
framework to implement this stochastic sampling in a non--local
fashion while maintaining detailed balance. CTDMC is well suited to
finding exact ground--state properties, applicable to any lattice
model not suffering from the sign problem; for a lattice model the
evolution of the wave function can be performed in continuous time
without any time discretization error. Both the directed loop
algorithm and the CTDMC are important recent advances in development
of computational methods. Here we present results for a Hubbard model for
anti--ferromagnetic spin--1 bosons in one dimensions, and show
evidence for a dimerized ground state in the lowest Mott lobe.
\end{abstract}

\maketitle

\section{Introduction}

An optical lattice can be made by applying orthogonal laser beams to
an ultracold gas of atoms. As a result $^{87}$Rb or $^{23}$Na atoms
can be trapped to form a perfect lattice. These atoms have a hyperfine
spin 1, with either a ferromagnetic ($^{87}$Rb) or antiferromagnetic
($^{23}$Na) interaction. Unpolarized $^{23}$Na atoms have
spin--correlated Mott insulating states \cite{demler-zhou-02}, and it
has been suggested \cite{yip-03} that the ground state is a dimer
phase in one, two and three dimensions. An effective Hamiltonian of
spin--1 bosons in an optical lattice has the Bose--Hubbard form,
supplemented with a term that describes the spin interaction of
atoms on the same lattice site \cite{imambekov-lukin-demler-04},
\begin{eqnarray}
\hspace*{-6pt}H =  -t\sum_{\langle ij\rangle,\alpha} \left(a_{i\alpha}^\dagger a_{j\alpha} 
+ cc. \right) 
-\mu\sum_i  \hat n_i + \frac{U}{2}\sum_i \hat n_i(\hat n_i-1)
 + \frac{U_2}{2}\sum_i \left({\vec S_i}^2 - 2\hat n_i\right)
\label{eq:ham} ,
\end{eqnarray}
where the spin components $\alpha=x,y,z$ form a basis where the spin
operator at site $i$ is written in terms of boson operators
$a^\dagger_\alpha$ as $S_i^\alpha = -i \epsilon^{\alpha \beta \gamma}
a^\dagger_\beta a_\gamma$.  ($\epsilon^{\alpha \beta \gamma}$ is the
totally antisymmetric Levi-Civita tensor). In this basis the
antiferromagnetic spin interaction has no sign problem.  The value of
$U_2/U$ can be determined from experimental scattering lengths.
 
SSE \cite{sandvik-kurkijarvi-91} is an exact scheme, where the quantum
partition function is expanded as a power series in a given basis
$\{|s\rangle\}$,
\begin{equation}
Z = {\rm Tr} \left\{e^{-\beta H} \right\} = \sum_{s}\sum_{n=0}^\infty
\frac{-\beta^n}{n!} \langle s| H^n |s\rangle\ ,
\label{eq:Z}
\end{equation}
where $\beta = 1/(k_BT)$ is the inverse temperature. Typically the
low--temperature phases we are simulating have at most 1--3 atoms per
site, so even with the spin degrees of freedom the required number of
single--site states is rather small. 
The Hamiltonian (\ref{eq:ham}) couples at most two
neighboring sites, so SSE splits the
partition function to a  sum of  products of bond and site operations.

To generate the terms in SSE we employ directed loop
updates\cite{syljuasen-sandvik-02}. The basic idea is to pick a few
well chosen elementary update operations, that change the bond or site
operators to each other. Each update affects only a single site (and
operator) at a time and is applied in a loop among the operators in
the product that is the current term of the SSE sum. The outcome is a
new product of operators, and, thanks to the looping, this new product
is also a term in the partition function.  In addition one controls
the number of operators in the product by adding or removing diagonal
operators; they don't change the state so their addition or removal
won't disrupt the continuity of the sequence. The rules for how the
loop travels follow from the detailed balance condition, which is used
to ascertain that the operators appear in the products with proper weights.

If the Hilbert space is finite one can formulate Diffusion Monte Carlo
(DMC) in continuous time \cite{farhi-gutmann-92,syljuasen-05b}.  DMC
is a stochastic simulation of the imaginary-time evolution operator
$e^{-H \tau}$. For an infinitesimal time step $d \tau$ this operator
describes one out of three possible actions: 1) No action, the current
state is unchanged, 2) Transition to a new state, and 3) The weight of
the current state changes. Action 3) is needed due to the in general
non-Markovian nature of the evolution operator.  The probabilities for
the different actions $p(i)$, $i=1,2,3$ can be read directly off the
Hamiltonian. The probability $p(2)$ is determined by off-diagonal
elements of the Hamiltonian and is therefore of order $d\tau$. Also
$p(3)$ is of order $d\tau$ because it describes the deviation from
Markovian evolution, thus only $p(1)$ is of order unity. It follows
that the DMC can be treated as a continuous--time simulation of a
radioactive (multi-channel) decay problem, where decay--times are
generated according to the exponential distribution: $\tau_{\rm
decay}= -\ln(r) d\tau/(1-p(1))$ where $r$ is a random number uniformly
distributed between 0 and 1. Having generated a decay--time the system
is moved directly to the time of decay and the appropriate decay
process is chosen dependent on the ratio $p(2)/p(3)$.  CTDMC can be
combined with other standard improvements of DMC such as importance
sampling, reweighting and
forward-walking\cite{nightingale-umrigar-98}.

\section{Results}

\begin{figure}[th]
\hbox{
\centerline{\includegraphics[width=0.4\textwidth]{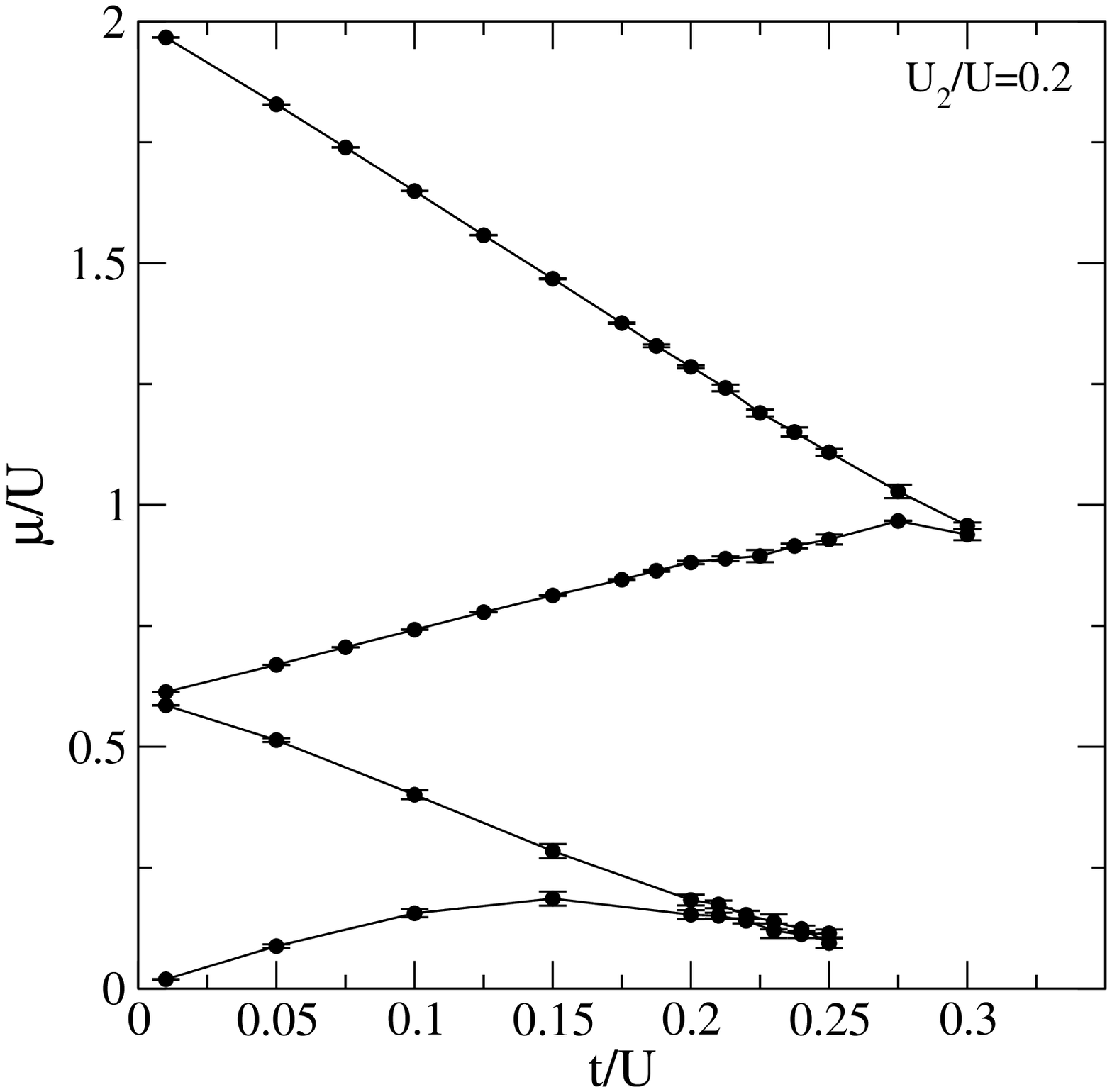}
\includegraphics[width=0.4\textwidth]{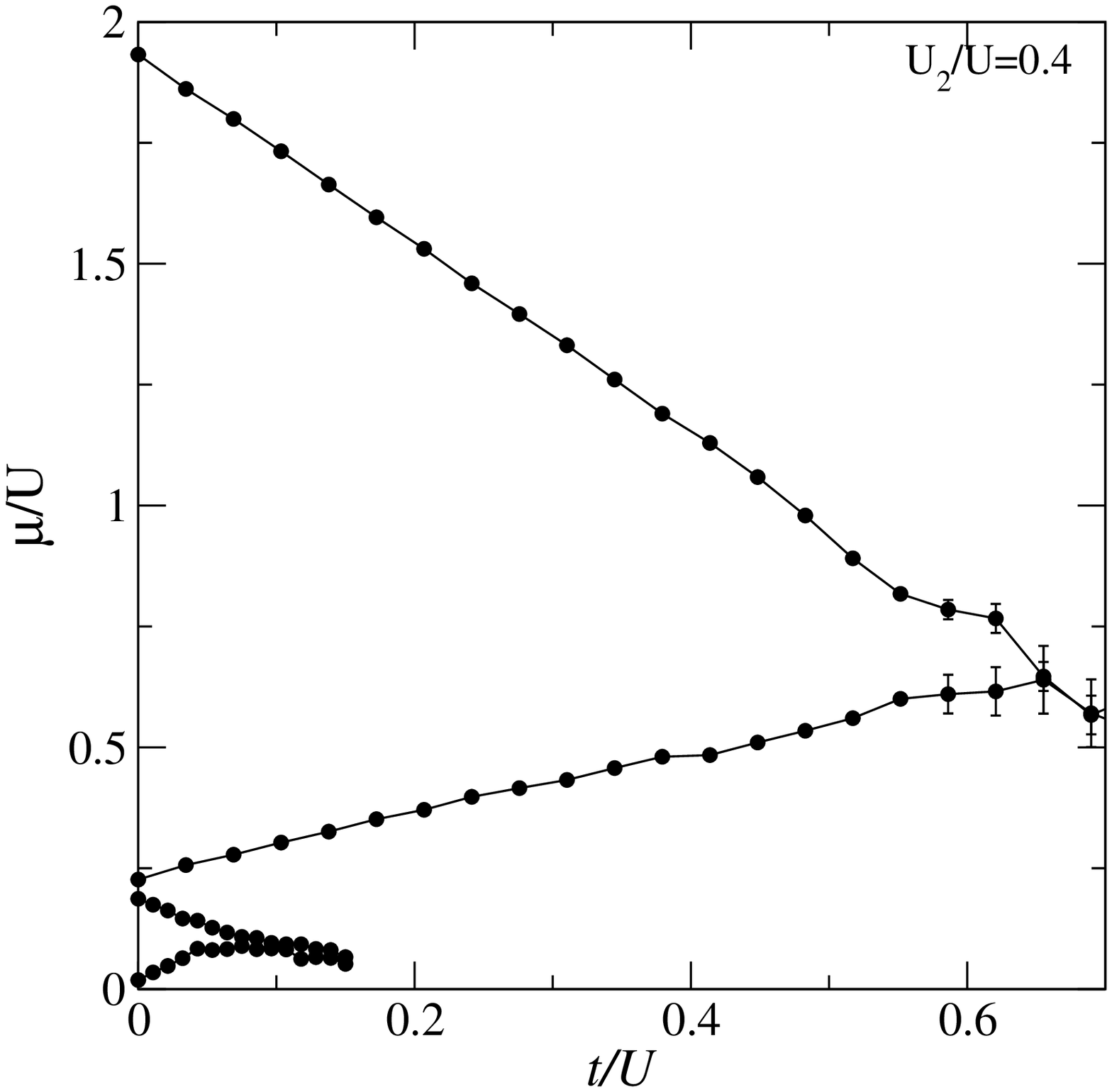}}
}
\vspace*{8pt}
\caption{The phase boundaries between Mott insulating and superfluid
regions of a 1D chain for $U_2/U=0.2$ (left panel) and $U_2/U=0.4$
(right panel) at zero temperature. The results were obtained using
CTDMC to compute ground state energies $E_L(N)$ for different particle
numbers $N$ and system sizes $L$. The upper(lower) boundary
$\mu_{p+}$($\mu_{p-}$) of the Mott lobe with $p$ bosons per site was
obtained from the finite size values for $\mu_{p\pm} = \pm \left(
E_L(pL\pm 1)-E_L(pL) \right)$ extrapolated to $L \to \infty$ using
$L=8,16,20$ and $24$.}
\label{fig:1mott}
\end{figure}

Fig.~\ref{fig:1mott} shows the phase boundaries of the two lowest Mott
insulating phases in the cases $U_2/U=0.2$ and $U_2/U=0.4$. Scattering
between spin states stabilizes the second lobe, while the first lobe
is slightly reduced in size. 

\begin{figure}[th]
\centerline{
\includegraphics[width=0.5\textwidth]{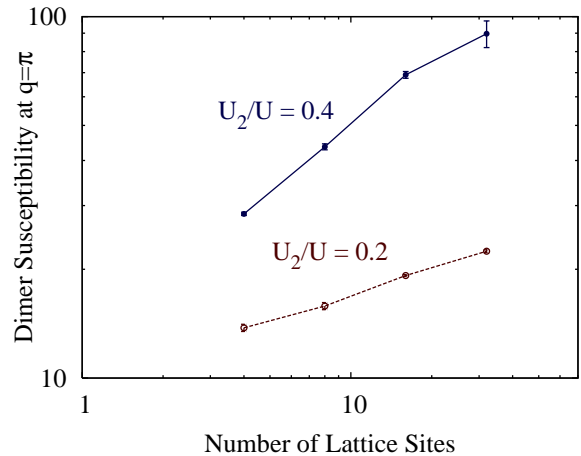}
}
\vspace*{8pt}
\caption{Finite size scaling of the 
dimer susceptibility ${\cal X}(q)$ at $q=\pi$ in the Mott
insulating phase of a 1D chain at $\beta=150$ for two values of $U_2/U$ indicated in the figure.
We show the results for $L=$ 4, 8, 16 and 32 sites, those 
for $U_2/U=0.4$ were computed at $t=0.1$, $\mu=0.1$, and 
those for $U_2/U=0.2$ at $t=0.15$, $\mu=0.25$. 
}
\label{fig:2susc}
\end{figure}

To study the structure of the Mott insulating phase with one atom per
site we have measured the bond hopping or dimer susceptibility
\begin{equation}
{\cal X}(q) = \frac{1}{t^2 \beta N} (\langle  N_q N_{-q}\rangle  
- \langle N_q\rangle\langle N_{-q}\rangle )
\label{eq:X}\ ,
\end{equation}
where $N_q$ is the Fourier transform of $N_i$, the number of hopping
operators on bond $i$. Because of the on--site spin scattering, the
hopping operators tend to couple pairwise adjacent sites, indicated by
a peak at wave number $q=\pi$.  For a dimer state in an infinite
system this peak height should diverge, and we demonstrate this in
Fig.~\ref{fig:2susc}. For larger than 32 sites the computation of the
susceptibility becomes exceedingly slow, as autocorrelation times
increase rapidly.  This is due to our non-optimal treatment of the
terms of type $a^\dagger_x a^\dagger_x a_y a_y$ which changes two spin
indices on the same site simultaneously. In order to build such terms
out of loop changes where only one spin index changes we have included
intermediate, auxiliary terms $a^\dagger_x a_y$ in the Hamiltonian. These auxiliary
terms are generally present when the loop is being built, but we do
not allow a loop to close if any of them  are present. 

\begin{figure}[th]
\hbox{
\centerline{
\includegraphics[width=0.5\textwidth]{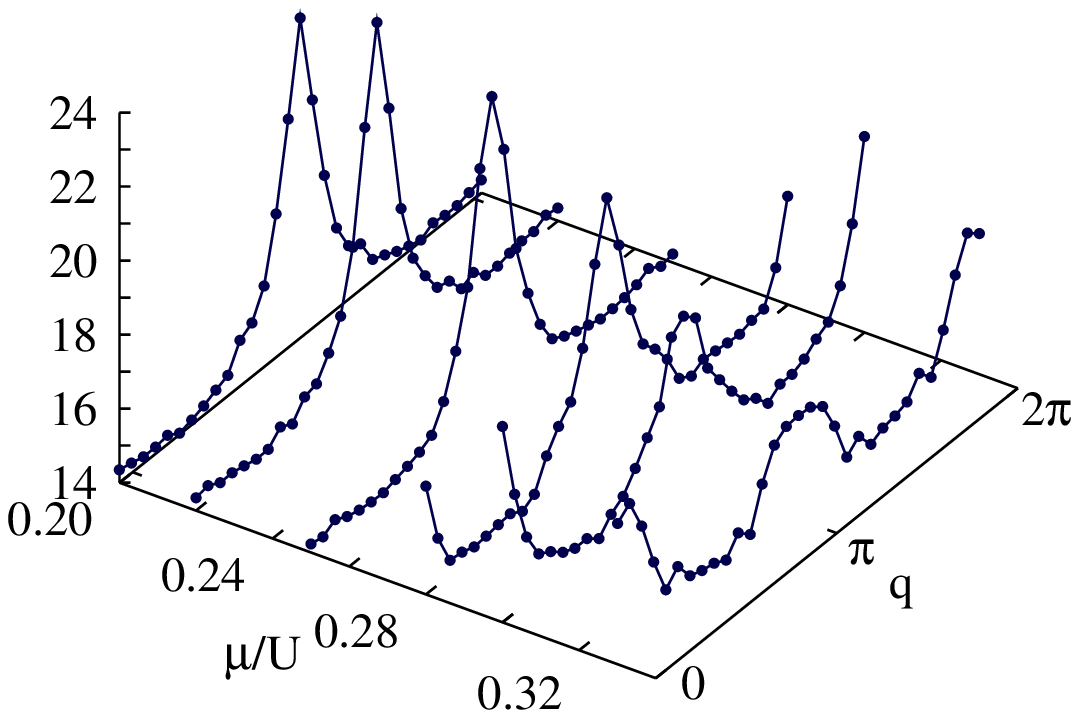}
\includegraphics[width=0.5\textwidth]{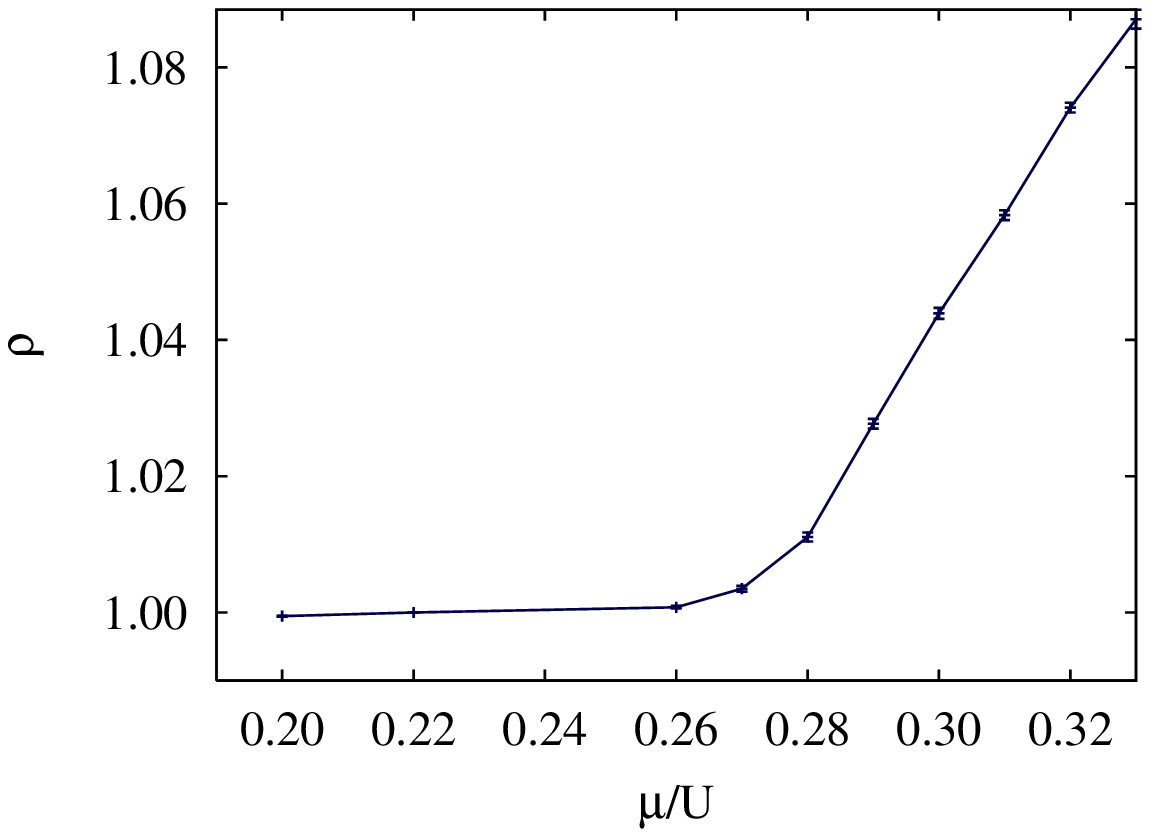}}
}
\vspace*{8pt}
\caption{The left panel shows the dimer susceptibility ${\cal X}(q)$
at the upper edge of the first Mott lobe (see the left panel of
Fig.~1) of a 1D chain with 32 sites.  The susceptibility is plotted as
a function of the chemical potential $\mu/U$ and the wave number $q$.
The data was computed at $t=0.15$, $\beta=150$ and $U_2/U=0.2$.  The
statistical error is less than 0.2 in the vertical scale.  The right
panel shows the corresponding density.  }
\label{fig:3susc}
\end{figure}

The left panel of Fig.~\ref{fig:3susc} show how the dimer state is suppressed
as one moves from the insulating to the superfluid state. The hopping
parameter $t$ is kept fixed and we move out of the insulating phase by
increasing $\mu$. The corresponding density, as it increases from
unity, is shown in the right panel of Fig.~\ref{fig:3susc}.


\begin{thebibliography}{1}

\bibitem{demler-zhou-02}
E. Demler and F. Zhou, Phys. Rev. Lett. {\bf 88},  163001  (2002).

\bibitem{yip-03}
S.~K. Yip, Phys. Rev. Lett. {\bf 90},  250402  (2003).

\bibitem{imambekov-lukin-demler-04}
A. Imambekov, M. Lukin, and E. Demler, Phys. Rev. Lett. {\bf 93},  120405
  (2004).

\bibitem{sandvik-kurkijarvi-91}
A.~W. Sandvik and J. Kurkij\"arvi, Phys. Rev. B {\bf 43},  5950  (1991).

\bibitem{syljuasen-sandvik-02}
O.~F. Sylju{\aa}sen and A.~W. Sandvik, Phys. Rev. E {\bf 66},  046701  (2002).

\bibitem{farhi-gutmann-92}
E. Farhi and S. Gutmann, Ann. Phys. {\bf 213},  182  (1992).

\bibitem{syljuasen-05b}
O.~F. Sylju{\aa}sen, Phys. Rev. B {\bf 71},  020401(R)  (2005).

\bibitem{nightingale-umrigar-98}
M.~P. Nightingale and C.~J. Umrigar, in {\em Advances in Chemical Physics} {\bf 105}, eds. D.~M. Ferguson {\em et al}. (John Wiley, NY, 1998), chapter 4.

\end{thebibliography}
\end{document}